\documentclass{iopart}

\usepackage{graphicx}

\def\simge{\mathrel{
       \rlap{\raise 0.511ex \hbox{$>$}}{\lower 0.511ex \hbox{$\sim$}}}}
\def\simle{\mathrel{
       \rlap{\raise 0.511ex \hbox{$<$}}{\lower 0.511ex \hbox{$\sim$}}}}

\begin{document}

\title[Thermodynamics of two-flavor lattice QCD 
with an improved Wilson quark action
]
{Thermodynamics of two-flavor lattice QCD 
with an improved Wilson quark action at non-zero 
temperature and density
}

\author{Y.~Maezawa$^1$\footnote{speaker},
S.~Aoki$^2$, S.~Ejiri$^3$, T.~Hatsuda$^1$, 
N.~Ishii$^4$, K.~Kanaya$^2$ and N.~Ukita$^4$}

\address{$^1$Department of Physics, The University of Tokyo, 
Tokyo 113-0033, Japan}
\address{$^2$Graduate School of Pure and Applied Sciences, 
University of Tsukuba, Tsukuba, Ibaraki 305-8571, Japan}
\address{$^3$Physics Department, Brookhaven National Laboratory,
Upton, New York 11973, USA}
\address{$^4$Center for Computational Sciences,
University of Tsukuba, Tsukuba, Ibaraki 305-8571, Japan}
\ead{maezawa@nt.phys.s.u-tokyo.ac.jp}

\begin{abstract}
We report the current status of our systematic studies of the QCD thermodynamics 
by lattice QCD simulations with two flavors of improved Wilson quarks. 
We evaluate the critical temperature of two flavor QCD in the chiral limit at zero chemical potential and show the preliminary result. Also we discuss
fluctuations at none-zero temperature and density 
 by calculating the quark number and isospin susceptibilities and their
  derivatives with respect to chemical potential.
\end{abstract}


\section{Introduction}

In order to extract unambiguous signals for the QCD phase transition from 
the heavy-ion collision experiments, quantitative calculations from first 
principle are indispensable. 
At present, the lattice QCD simulation is the only systematic method 
to do so and various interesting results have been already reported. 
So far, most lattice QCD studies at finite temperature $(T)$ 
and chemical potential $(\mu_q)$ have been performed using staggered-type 
quark actions with the fourth-root trick of the quark determinant.
Thus, the other actions such as the Wilson-type quark actions are
 necessary  to control and estimate systematic errors due to 
 lattice discretization.

Several years ago, the CP-PACS Collaboration has studied the QCD 
thermodynamics using the Iwasaki (RG) improved gauge action and the 
two-flavor clover improved Wilson quark action \cite{cp1}. 
We revisit this action armed with recent techniques 
at finite chemical potential $\mu_q$.
In this report, we present on our preliminary results 
for the critical temperature at $\mu_q=0$
as well as the quark number and isospin susceptibilities at finite $\mu_q$. 
The latter are related to the physics  of the possible critical point 
in the $(T,\mu_q)$ plane.

\section{Critical temperature of two-flavor QCD}

The critical temperature ($T_c$) is one of the most fundamental quantities 
in the QCD thermodynamics.
We renew the analysis of $T_c$ done in \cite{cp1}
performing additional simulations at $\beta=6/g^2=1.8$ for $N_t=4$ 
and $1.95$ for $N_t=6$.
We determine the pseudo-critical point $\beta_{pc}$ defined from 
 the peak of Polyakov loop susceptibility on 
 $N_s^3 \times N_t = 16^3 \times 4$ and $16^3 \times 6$ lattices, 
 as a function of the hopping parameter $K$.

\begin{figure}[tb]
  \begin{center}
    \begin{tabular}{cc}
      \includegraphics[width=57mm]{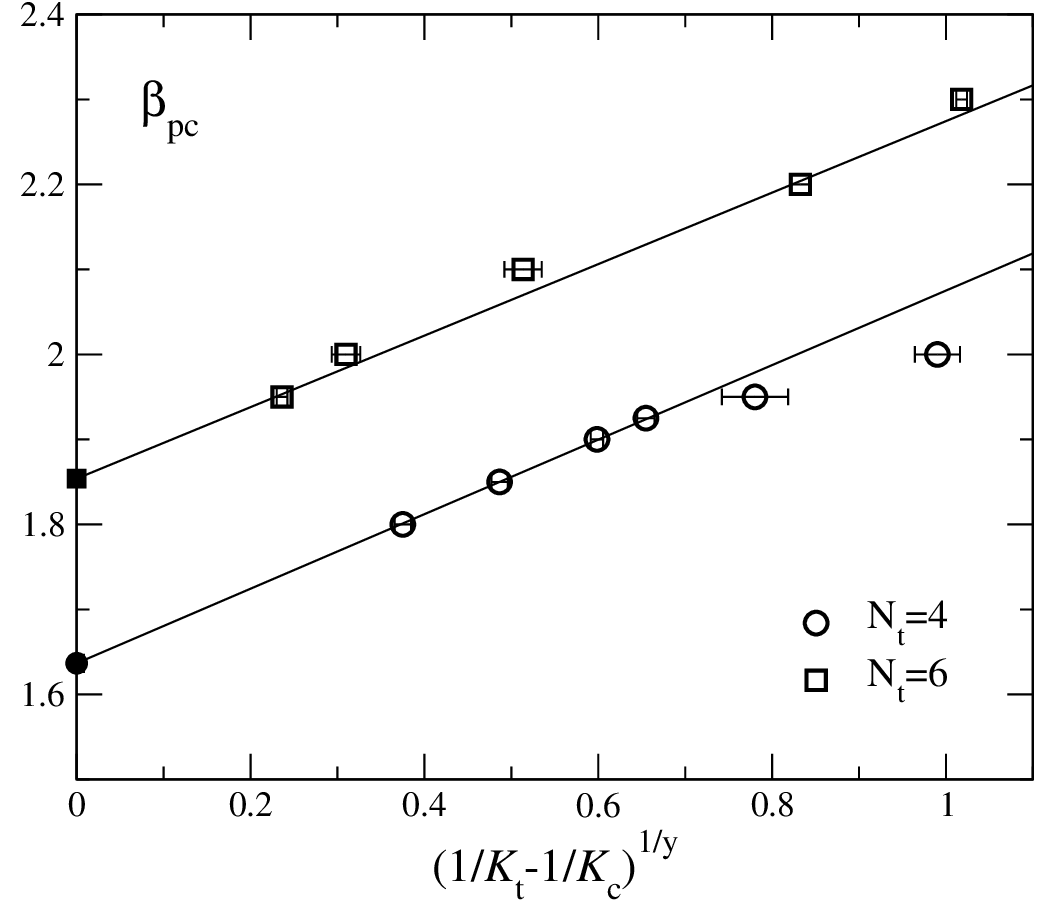} &
      \includegraphics[width=57mm]{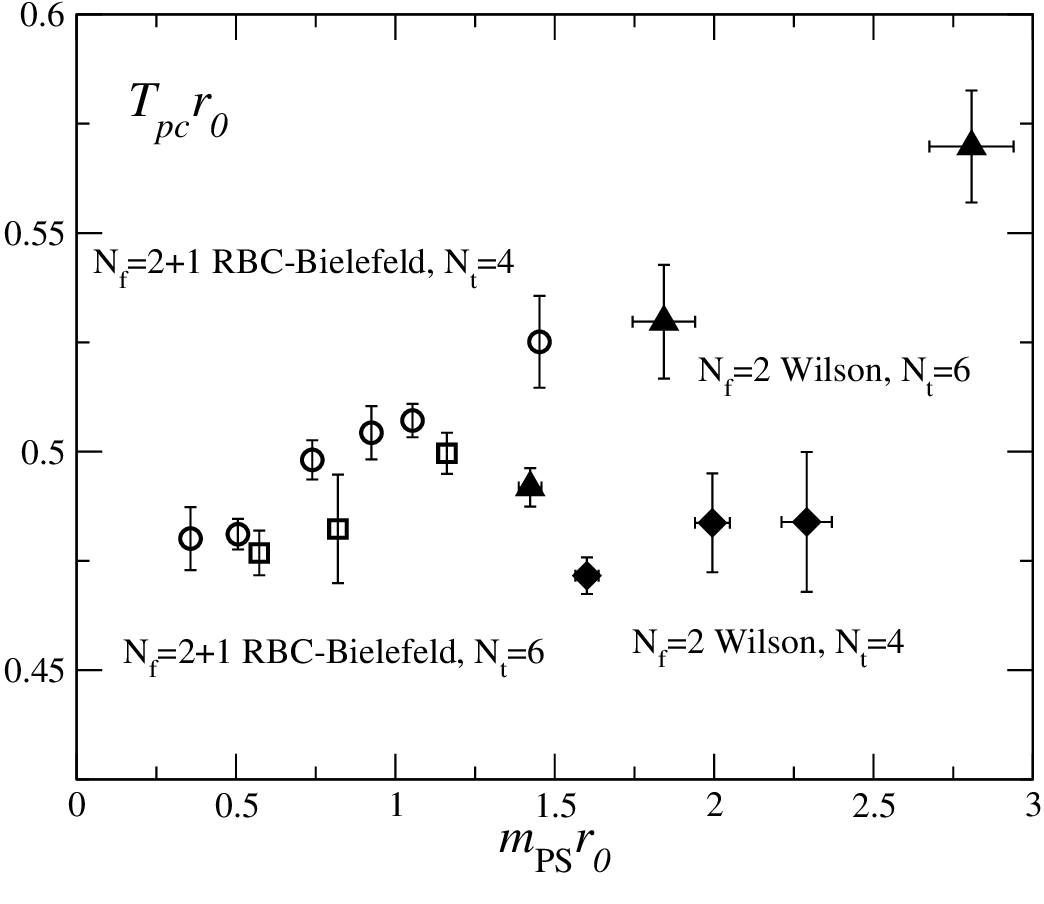}
    \end{tabular}
    \vspace*{-2mm}
    \caption{Left: The pseudo-critical point $\beta_{pc}$
    as a function of $m_q \sim 1/K - 1/K_c$ for $N_t=4$ (circle) 
    and $N_t=6$ (square).  
    Right: Comparison of $T_{pc}$ scaled by $r_0$ between 
    the staggered quark action (open symbol) and the Wilson quark action
    (filled symbol) for $N_t=4$ and 6.
    }
    \label{fig1}
  \end{center}
\vspace{-0.5cm}
\end{figure}

 It is confirmed in Ref.~\cite{cp1} that a subtracted chiral condensate 
 satisfies the scaling behavior with the critical exponents and 
 scaling function of the three-dimensional O(4) spin model.
 The reduced temperature $t$ and external magnetic field $h$ are 
 identified to $t \sim \beta-\beta_{ct}$ and $h \sim m_q$, respectively, 
 where $\beta_{ct}$ is the critical transition point in the chiral limit.
 Assuming that the pseudo-critical temperature
from the Polyakov loop susceptibility
  follows the same scaling
law as the O(4) spin model, $t_{pc} \sim h^y$
with $y = 0.537(7)$, 
 we fit the data of $\beta_{pc}$ by
 $\beta_{pc}=\beta_{ct}+Ah^{1/y}$
  with two free parameters, $\beta_{ct}$ and $A$, in the range of 
$\beta=1.8$--1.95 for $N_t=4$ and $\beta=1.95$--2.10 for $N_t=6$. 
First, we adopt the definition $m_q a \sim 1/K - 1/K_c$ as the quark mass
where $K_c$ is the chiral point where the pion mass vanishes
 at $T=0$ for each $\beta$. 
See Fig.~\ref{fig1} (left). 
We calculate the critical temperature $T_c$ in the chiral limit 
using $T=1/N_t a$. The lattice spacing $a$ is estimated from the vector 
meson mass assuming $m_{\rm V}(T=0)=m_{\rho}=770$ MeV at $\beta_{ct}$ on $K_c$. 
By this procedure,
we obtain the preliminary results $T_c = 183(3)$ MeV for $N_t=4$ and 
174(5) MeV for $N_t=6$. 
We also calculate $\beta_{ct}$ using the relation of 
$m_q^{\rm AWI} \propto m_{\rm PS}^2$, 
where $m_{\rm PS}$ is the pseudo-scalar meson mass  and $m_q^{\rm AWI}$ is
 the quark mass obtained from the axial vector Ward-Takahashi identity. 
The results of $T_c$ are 173(3) MeV $(N_t=4)$, 167(3) MeV $(N_t=6)$ for 
$h=(m_{\rm PS} a)^2$ and 176(3) MeV $(N_t=4)$ for $h=m_q^{\rm AWI} a$.
We note that these O(4) fits reproduce the data of $\beta_{pc}$ much better 
than a linear fit $\beta_{pc}=\beta_{ct}+Ah$. 
From these analyses, we tentatively conclude that
the critical temperature in the chiral limit is in 
the range 170--186 MeV for $N_t=4$ and 164--179 MeV for $N_t=6$. 
There is still a large uncertainty  from the choice of the fit ansatz. 
To remove this, we are performing further simulations at lighter quark masses.

Next, we compare our results with those of the staggered quark action.
We plot the results of the pseudo-critical temperature ($T_{pc}$) 
in unit of Sommer scale $(r_0)$ as a 
function of $m_{\rm PS} r_0$ in Fig.~\ref{fig1} (right) together with 
those by the RBC-Bielefeld Collaboration using 2+1 flavor p4-improved staggered 
quark action \cite{Cheng:2006qk}. 
As seen in this figure, results of $T_{pc}$ obtained by different quark 
actions seem to approach the same function of $m_{\rm PS} r_0$ 
as $N_t$ increases.

A care is in order when 
we convert  $T_c r_0$ ($T_{pc} r_0$) to $T_c$ ($T_{pc}$)
 in MeV using a physical value of $r_0$. 
Because the phenomenological estimate of $r_0$ has large theoretical uncertainties, 
it looks convenient to adopt a lattice result.
Unfortunately, lattice results for $r_0$ suffer from sizable ambiguities yet.
The RBC-Bielefeld Collaboration adopted the value $r_0 = 0.469(7)$ fm, 
which was obtained by Gray {\it et al.} \cite{Gray:2005ur} from a bottomonium mass splitting using the AsqTad-improved staggered quark action.
On the other hand, 
the CP-PACS+JLQCD Collaboration found $r_0=0.516(21)$ fm from a study 
of light hadron spectrum 
using the clover-improved Wilson quark action \cite{Ishikawa:2006ws}. 
This leads to about 10 \%
 difference in the value of $T_c$ and makes it difficult 
 to naively compare results from different groups.

\section{Hadronic fluctuations at finite $\mu_q$}

 Hadronic fluctuations at finite density are observables closely related to 
the critical point in the $(T, \mu_q)$ plane and may be experimentally 
detected by an event-by-event analysis of heavy ion collisions. 
 The fluctuations can also be studied by numerical simulations of 
lattice QCD calculating the quark number and isospin susceptibilities,  
$\chi_q$ and $\chi_I$. They correspond to
 the second derivatives of the pressure with respect to $\mu_q$ 
 and $\mu_I$, where $\mu_I$ is the isospin chemical potential. 
 From a phenomenological argument in the sigma model, 
 $\chi_q$ is singular at the critical point, 
 whereas $\chi_I$ shows no singularity there. 
 
We perform simulations at $m_{\rm PS} / m_{\rm V} = 0.65$ and 0.80  
on a $16^3 \times 4$ lattice.
We calculate the $\chi_q$ and $\chi_I$ and their second derivatives  
with respect to $\mu_q$ and $\mu_I$ 
at $\mu_q = \mu_I =0$. (Note that the odd derivatives are zero at $\mu_q=0$.) 
The details of the calculations are reported in Ref.~\cite{ejiri}.

\begin{figure}[tb]
  \begin{center}
    \begin{tabular}{cc}
      \includegraphics[width=57mm]{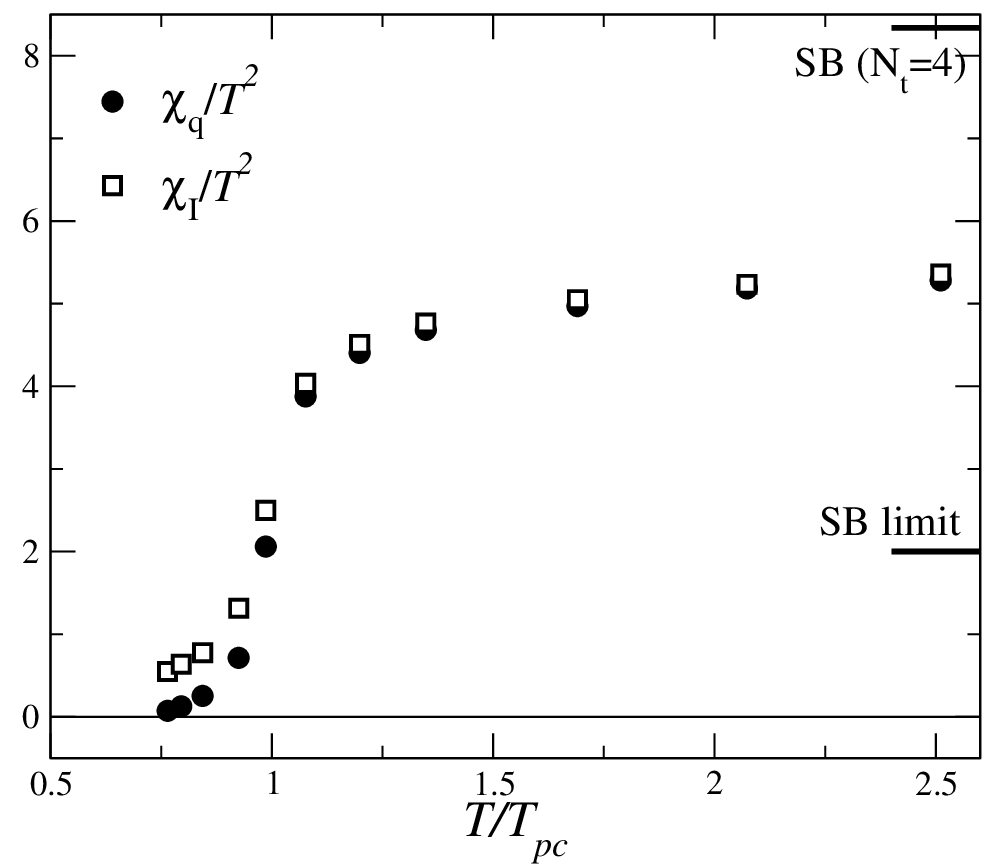} &
      \includegraphics[width=57mm]{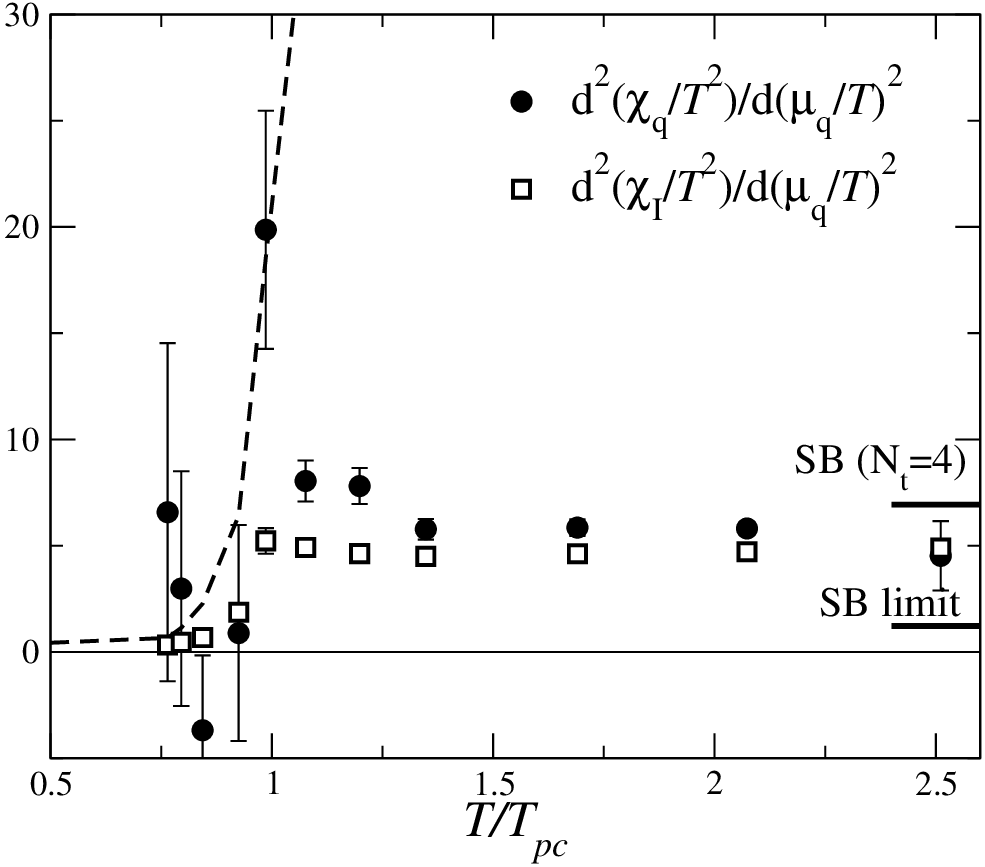} 
    \end{tabular}
    \vspace*{-2mm}
    \caption{Left: Quark number (circle) and isospin (square)
    susceptibilities at $\mu_q = \mu_I = 0$.
    Right: The second derivatives of these susceptibilities.
    }
    \label{fig2}
  \end{center}
\vspace{-0.5cm}
\end{figure}

The left panel of Fig.~\ref{fig2} shows 
 $\chi_q/T^2$ (circle) and $\chi_I/T^2$ (square) 
 at $m_{\rm PS}/m_{\rm V}=0.8$ and $\mu_q = \mu_I =0$ as functions of $T/T_{pc}$.
We find that $\chi_q/T^2$ and $\chi_I/T^2$ 
increase sharply at $T_{pc}$, in accordance with the expectation 
that the fluctuations in the QGP phase are much larger 
than those in the hadron phase.
Their second derivatives $\partial^2 (\chi_q/T^2)/ \partial (\mu_q/T)^2$ 
and $\partial^2 (\chi_I/T^2)/ \partial (\mu_q/T)^2$ are shown 
in Fig.~\ref{fig2} (right). 
We find that basic features are quite similar to those found previously with the  p4-improved staggered fermions \cite{BS03}. 
$\partial^2 (\chi_I/T^2)/ \partial (\mu_q/T)^2$ remains small around $T_{pc}$,
suggesting that there are no singularities in $\chi_I$ at non-zero density. 
On the other hand, we expect a large enhancement in the quark number 
fluctuations near $T_{pc}$ as approaching the critical point in the 
$(T, \mu_q)$ plane. 
The dashed line in Fig.~\ref{fig2} (right) 
is a prediction from the hadron resonance gas model,
  $\partial^2 \chi_q/ \partial \mu_q^2 \approx 9 \chi_q/T^2$.
Although  current statistical errors 
in Fig.~\ref{fig2} (right) are still large, 
we find that  $\partial^2(\chi_q/T^2) / \partial (\mu_q/T)^2$ near $T_{pc}$ is 
much larger than that at high temperatures.
At the right end of the figure, values 
of free quark-gluon gas (Stefan-Boltzmann gas) 
 for $N_t=4$ and for $N_t = \infty$ limit are shown.
 Since the lattice discretization error in the equation of state is 
 known to be large at $N_t=4$ with our quark action,
 we need to extend our study to larger $N_t$ for the continuum extrapolation.

\section{Conclusion}

We reported the current status of our study of QCD thermodynamics 
using the two-flavor improved Wilson quark action.
The critical temperature is estimated from the Polyakov loop susceptibilities
on $16^3 \times 4$ and $16^3 \times 6$ lattices.
We discussed uncertainties from the chiral extrapolation and the scale parameter.
Our preliminary result of $T_c$ in the chiral limit is in the range of 
$T_c=170$--186 MeV for $N_t=4$ and $T_c=164$--179 MeV for $N_t=6$.

The fluctuations of quark number and isospin densities were also discussed.
Although the statistical errors are still large, we find that 
$\chi_q$ seems to increase rapidly near $T_{pc}$ as $\mu_q$ increases, 
whereas the increase of $\chi_I$ is not large near $T_{pc}$. 
These behaviors qualitatively agree with the previous 
results obtained with the p4-improved staggered fermions.

We also studied the static quark free energies,
Debye screening masses and spatial string tension at finite
temperature with the same action, which are reported in detail in Ref.~\cite{FE}.

\ack
This work is in part supported by Grants-in-Aid of the Japanese 
MEXT
(Nos.~13135204, 15540251, 17340066, 18540253, 18740134).
SE is supported by Sumitomo Foundation (No.~050408), and 
YM is supported by JSPS.
This work is in part supported also by ACCC, Univ. of Tsukuba, 
and the Large Scale Simulation Program No.06-19 (FY2006) of
 KEK.

\end{document}